# Density functional theory for fractional charge: Locality, size consistency, and exchange-correlation

Jing Kong*

Department of Chemistry and Center for Computational and Data Sciences, Middle Tennessee State University, 1301 Main St., Murfreesboro, TN 37130, USA

* Correspondence: jing.kong@mtsu.edu

Abstract

We show that the exact universal density functional of integer electronic charge leads to an extension to fractional charge in an asymptotic sense when it is applied to a system made of asymptotically separated densities. The extended functional is asymptotically local and is said to be $i$-local. The concept of $i$-locality is also applicable to nuclear external potentials, and a natural association exists between the localities of a density and a set of nuclei. Applying the functional to a system with nuclei distributed in two asymptotically separated locales requires an explicit search of the electronic charge at each locale with the constraint of the global charge. The determined number of electrons at each locale can be fractional. The molecular size consistency principle is realized as the result of the search. It is physically sensible to extend the molecule concept to include a fractional number of electrons (called fractional molecule henceforth) as a localized observable. The physical validity of fractional molecules is equivalent to the asymptotic separability of molecules, a basic assumption in molecular research. A one-to-one mapping between a fractional molecule's density and external potential is shown to exist with a nondegenerate condition. The global one-to-one mapping required by the Hohenberg-Kohn first theorem is realized through the aforementioned global search for molecular charges. Furthermore, the well-known piecewise linearity of the universal functional with respect to the number of electrons is necessary for an approximate $i$-local universal functional to be broadly accurate for any integer number of electrons. The Kohn-Sham (KS) noninteracting kinetic energy functional for a fractional molecule is well defined and has the same form as that for a system of an integer number of electrons. It is shown to be $i$-local. A nondegenerate, noninteracting ensemble $v$-representable fractional density is simultaneously noninteracting wavefunction representable. A constrained search over those representing wavefunctions yields the definition of an exchange-correlation functional pertaining to fractional occupancies of KS orbitals. The functional is shown to be an upper bound to the formal Kohn-Sham exchange-correlation energy of a fractional molecule and includes a strong correlation. It yields the correct result for a well-designed example of effective fractional occupancies in the literature.

List of frequently used terms, their indexes of reference, and the page numbers with their explanations.

| Term | Index | Page |
| --- | --- | --- |
| fractional density | D1 | 4 |
| effective separation | D2 | 4 |
| asymptotic separation | D3 | 4 |
| AIND, ANDI | D4,C1 | 4 |
| molecular energy convexity | C2 | 7 |
| $i$-local, $i$-domain, $i$-locality association | D5,D6, D7 | 7 |
| fractional molecule | D8 | 9 |
| nondegeneracy of fractional molecule | C3 | 11 |
| auxiliary atom | D9 | 15 |
| reference molecule | D10 | 15 |

List of some frequently used symbols and the page numbers with their explanations.

| Term | Page |
| --- | --- |
| $\rho_{N_X}^{(X)}$ | 4 |
| $\rho_{N_X.N_Y}^{(X).(Y)}$ | 5 |
| $\rho_{N_X..N_Y}^{(X)..(Y)}$ | 5 |
| $\Psi_{\rho_L}(L)$ | 5 |
| $\cong$ | 5 |
| $\delta$ | 7 |
| $X_N$ | 9 |
| $\rho_N$ | 9 |
| $A_\delta$ | 16 |
| $\tilde{X}_N$ | 16 |

## Introduction

Density functional theory (DFT)[1], especially Kohn-Sham (KS) DFT with Slater determinants (SDs)[2], is the bread-and-butter method for simulating electronic effects in molecules and materials due to its efficiency and accuracy. Contemporary DFT methods, however, are inadequate or even fail for systems with significant nondynamic correlation or charge delocalization. The problem is most pronounced in a system where the local electronic charge (henceforth referred to as charge) or spin becomes effectively fractional, such as a monomer in an infinitely stretched $H_2^+$ (fractional charge) or $H_2$ (fractional spin)[3-5]. Both fractional charge and spin can happen simultaneously, such as with the dissociation of a hetero-diatomic molecule[6]. Large fractional charge and fractional spin errors with contemporary functionals are well-documented[7], and functionals that performed well for one type of error were found to do poorly for the other[8]. Efforts have been made to design methods within the KS framework[9-12] or in the proximity of it[13, 14] to address effective fractional occupancies in a molecular environment and demonstrated significant improvements over contemporary mainstream functionals.

An early formal treatment of fractional charge was Janak theorem[15]. It stated that the energy of a Kohn-Sham orbital of a molecule is the derivative of the total energy with respect to the occupancy of that orbital if the latter was allowed to vary continuously. Perdew *et al.* later defined a molecular system with a fractional charge as a statistical average of molecules with fluctuating integer charges and showed that the energy of the system was piecewise linear between integer charges[3, 16]. They also showed that the KS exchange-correlation (XC) potential, the derivative of KS XC energy with respect to the electron density, experienced a constant shift at an integer charge due to the energy piecewise linearity, the well-known derivative discontinuity[3]. The lack of proper derivative discontinuity of contemporary functionals was considered the root cause of their ills in dealing with nondynamic correlation and charge delocalization. Yang *et al.* gave an alternative definition of a molecule with fractional charge as the average of an aggregation of many of the same molecules kept far away from each other[17, 18]. The number of identical molecules in the aggregate is determined so that the averaged number of electrons or the total spin on each molecule is the desired fraction. This arrangement achieved the aforementioned piecewise linearity of the energy of each molecule via a linear decomposition of the energy of the whole aggregate inferred from the size-consistency principle. The energy of each molecule was further assumed to be the functional of its own density only inferred from the size consistency argument. This wavefunction-based definition allowed a general concept of a fractional spin state of a molecule that becomes part of an ensemble of the degenerate wavefunctions of the aggregate. The energy of this fractional spin state was shown to be a constant, equal to the energy of that molecule with normal integer spin states[5]. The straight-line conditions of fractional charge and fractional spin formed a so-called flat plane condition[7]. Fractional charge can also be formed through Levy constrained search in Fock space[19]. In all cases, the fractional charge/spin in a density is represented by the fractional occupancy at the highest occupied molecular orbital(s) (HOMO(s)) of the KS noninteracting solution with orbitals below the HOMO(s) (called subHOMO(s) henceforth) fully occupied, i.e., following the Aufbau rule[15, 20].

Fundamental issues still exist in the DFT treatment of fractional charge. In particular, the rigor of the approaches mentioned above was questioned in two recent papers by Baerends[6, 21]. It was argued that DFT is based on the wavefunction solution of a single molecule with an integer charge. The system with fractional charge in ref [3] was defined as the statistical thermodynamic ensemble of an open system, not a single molecule[16]. An aggregate of many identical molecules in ref[17] could be treated as one supermolecule, but the energy of each molecule in the aggregate would be formally a functional of the density of the whole aggregate. The locality of the exact functional, i.e., being a functional of the density of one molecule only, could not be trivially inferred from the size consistency principle because the exact functional has been shown to exhibit global behaviors[16, 22-25]. Without proper justification of fractional charge, the functional of a density of fractional charge is undefined, and the piecewise linearity of DFT energy would become an arbitrary choice instead of an exact property. It follows that the derivative of the energy with respect to the charge could not be uniquely defined, and the validity of the Janak theorem as an exact condition of the KS solution was questionable. We add that the applicability of the KS SD scheme remains an open question to molecules of fractional charge since the fractional charge is an ensemble in nature and thus seems to preclude the applicability of the KS SD scheme. This is particularly important for cases where the correlation is strong and can only be described by multi-determinants with wavefunction methods.

Given the issues discussed above, this paper investigates the following questions: a) Whether the fractional charge is intrinsic to DFT in the sense that whether the exact universal functional of integer charge results in a functional of fractional charge; b) Whether the piecewise linearity is a necessary property of the exact universal functional; c) Whether the KS SD scheme is applicable to systems of fractional charge and d) Whether the major component of the XC can be extracted as an upper bound.

We note that this paper is focused on the electronic energies and densities of ground electronic states. We refer to a positive real number as being *fractional*, following the convention of DFT literature. '*Charge*' means the number of electrons for brevity unless noticed otherwise.

## Universal functional of fractional charge

Let $\rho_{N_X}^{(X)}$ be a density that integrates to a positive real number $N_X$ and tends to zero effectively outside a limited spatial range $X$, i.e., $\rho_{N_X}^{(X)}$ effectively has a compact support. Such a density is called *fractional density* henceforth for brevity [Definition D1]. The fractional densities of two separate spatial ranges are *effectively separated* [Definition D2] because their overlap is effectively zero. Let $\rho_L$ be any density that integrates to a positive integer $L$ ($N$ is used for nonnegative reals and $L$ for nonnegative integers throughout this paper.). $\rho_L$ is always made of one or more fractional densities due to its positive definiteness, i.e., we assume that $\rho_L$ effectively has a compact support. We use 'density' and 'fractional density' interchangeably and use the latter to emphasize the continuity of its charge. The sum of two effectively separated fractional densities is noted as $\rho_{N_X.N_Y}^{(X).(Y)} \equiv \rho_{N_X}^{(X)} + \rho_{N_Y}^{(Y)}$, with '.' as a symbol for effective separation. We say that two effectively separated fractional densities $\rho_{N_X}^{(X)}$ and $\rho_{N_Y}^{(Y)}$ are *asymptotically* separated [Definition D3] when the closest distance between any part of $\rho_{N_X}^{(X)}$ and any part of $\rho_{N_Y}^{(Y)}$ is larger than a number such that the inverse of this number is considered effectively zero subject to a positive threshold. This condition is called *asymptotically negligible distant interaction* (ANDI), henceforth [Condition C1]. The distance at the threshold is called *asymptotically interaction-negligible distance* (AIND) [Definition D4] henceforth. The AIND can be arbitrarily large. We denote the sum of those two asymptotically separated densities as $\rho_{N_X..N_Y}^{(X)..(Y)}$, i.e., using '..' to indicate an asymptotic separation.

Let $F_{UN}[L, \rho_L]$ be a *un*iversal functional with a given integer charge $L$. It is defined in the literature as the universal (kinetic and interelectronic) part of the molecular energy with a density matrix determined by the variational principle with the constraint of the given density $\rho_L$ [26-28]. We assume $F_{UN}[L,\rho_L]$ being the same as Lieb's density-matrix-based functional $F_{DM}$ [27]. We denote a density matrix constrained by the density $\rho_L$ as $D_{\rho_L}$. It is generally accepted that a wavefunction for a system made of two asymptotically separated subsystems of integer charge can be accurately written as an antisymmetric product of two wavefunctions at the two respective locales (A *locale* refers to a place in space where a local system situates), referred as left-right (L-R) wavefunction here. For a given density $\rho_{N_X..N_Y}^{(X)..(Y)}$, it is sufficient to limit the search space of density matrices based on wavefunctions of the L-R type such that the density at each locale is represented by a density matrix at the locale. We denote such a density matrix as $D^{\bullet\bullet}_{\rho_{L_X..L_Y}^{(X)..(Y)}}(L_X+L_Y)$ (A more detailed definition of this density matrix can be found in Appendix.). Let $U[D_\rho]\equiv Tr(D_\rho(T+V_{ee}))$ symbolize the universal part of the electronic energy for a density matrix $D_\rho$. The universal parts of those density matrices for $\rho_{N_X..N_Y}^{(X)..(Y)}$ have the following relation:

$$U[D_{\rho_{L_X..L_Y}^{(X)..(Y)}}(L_X+L_Y)] \overset{ANDI}{\cong} U[D^{\bullet\bullet}_{\rho_{L_X..L_Y}^{(X)..(Y)}}(L_X+L_Y)] \overset{ANDI}{\cong} U[D_{\rho_{L_X}^{(X)}}(L_X)]+U[D_{\rho_{L_Y}^{(Y)}}(L_Y)] \tag{1}$$

The symbol $\cong$ is used to indicate being equal within an arbitrarily small but nonzero tolerance of error or threshold. Each of the last two terms is a local partial trace as part of the energy generated from a global L-R wavefunction or density matrix. A detailed derivation of Eq.(1) can be found in the Appendix.

Eq.(1) leads to the following asymptotic linear decomposition of $F_{UN}[L,\rho_L]$ when the latter is applied to two asymptotically separated densities of integer charge:

$$\begin{aligned} &F_{UN}[L_X+L_Y,\rho_{L_X..L_Y}^{(X)..(Y)}]=\min_{D_{\rho_{L_X..L_Y}^{(X)..(Y)}}} U[D_{\rho_{L_X..L_Y}^{(X)..(Y)}}(L_X+L_Y)] \overset{ANDI}{\cong} \min_{D_{\rho_{L_X}^{(X).}}} U[D_{\rho_{L_X}^{(X)}}(L_X)]+\min_{D_{\rho_{L_X}^{(X).}}} U[D_{\rho_{L_X}^{(X)}}(L_X)] \\ &=F_{UN}[L_X,\rho_{L_X}^{(X)}]+F_{UN}[L_Y,\rho_{L_Y}^{(Y)}] \end{aligned} \tag{2}$$

We call this property $i$*-locality* ($i$ for interaction) [Definition D5]. $F_{UN}$ at each locale is a minimum energy for the density of an integer charge at the locale. More generally, let a functional $F[N,\rho_N]$ be called $i$*-local* if it satisfies the following relation with the ANDI condition:

$$F[N_X+N_Y,\rho_{N_X..N_Y}^{(X)..(Y)}]=F[N_X,\rho_{N_X}^{(X)}]+F[N_Y,\rho_{N_Y}^{(Y)}] \tag{3}$$

$I$-locality means that each of the asymptotically separated densities may be treated computationally separately.

It is logical to ask for the reason of completeness if and how a universal functional of fractional density can be defined that maintains the $i$-locality since a density of integer charge may contain asymptotically separated fractional pieces. We show that $F_{UN}[L,\rho_L]$ leads to the definition of $F_{UN}[N,\rho_N]$ with $N$ being real and positive when the former is applied to a density composed of two asymptotically separated pieces. For a density $\rho_{N_X..N_Y}^{(X)..(Y)}$ with $N_X+N_Y=L$, the electronic movement of this density is described by ensembles of density matrix $D_{\rho_{L_X..L-L_X}^{(X)..(Y)}}(L)$ with various $L_X$ for a given $L$. We call each of the fractional densities ensemble representable henceforth in the sense that it is derivable from a canonical ensemble even though the latter may include more electrons. We can decompose the universal functional of the sum of two asymptotically separated fractional densities into two local sides according to Eq.(1):

$$
\begin{aligned}
&F_{UN}[L,\rho^{(X)..(Y)}_{N..(L-N)}] \overset{ANDI}{\cong} \min_{D^{\cdot\cdot}_{\rho^{(X)..(Y)}_{i..(L-i)}},\delta_i\geq 0,\sum_i \delta_i=1} \sum_i \delta_i U[L,D^{\cdot\cdot}_{\rho^{(X)..(Y)}_{i..(L-i)}}] \\
&\overset{ANDI}{\cong} \min_{D_{\rho^{(X)}_i},D_{\rho^{(Y)}_{L-i}},\delta_i\geq 0,\sum_i \delta_i=1,\sum_i \delta_i\rho^{(X)}_i=\rho^{(X)}_N,\sum_i \delta_i\rho^{(Y)}_{L-i}=\rho^{(Y)}_{L-N}} \sum_i \delta_i (U[i,D_{\rho^{(X)}_i}]+U[L-i,D_{\rho^{(Y)}_{L-i}}]) \\
&= F_{UN}[N,\rho^{(X)}_N]+F_{UN}[L-N,\rho^{(Y)}_{L-N}]
\end{aligned}
\tag{4}
$$

with

$$
F_{UN}[N,\rho_N] \equiv \min_{\delta_i\geq 0,\sum_i \delta_i=1,\sum_i \delta_i\rho_i=\rho_N} \sum_i \delta_i F_{UN}[i,\rho_i] \tag{5}
$$

The extended $F_{UN}$ in Eq.(5) has the same form as the grand canonical universal functional in ref.[28] and was shown to be convex. Its convexity ensures its uniqueness. Its value at each locale is a minimum energy for the fractional density at the locale.

Eq.(4) is extendable to a system composed of many asymptotically separated fractional densities (See Eq.(50) in Appendix.). By the same argument, $F_{UN}[N,\rho_N]$ is $i$-local for a noninteger $N$ because the sum of two asymptotically separated fractional densities together forms a fractional density that can always be considered as part of a system of integer charge, i.e.,

$$
F_{UN}[N_X+N_Y,\rho^{(X)..(Y)}_{N_X..N_Y}]=F_{UN}[N_X,\rho^{(X)}_{N_X}]+F_{UN}[N_Y,\rho^{(Y)}_{N_Y}] \tag{6}
$$

We make some more observations about $F_{UN}[N,\rho_N]$ defined in Eq.(5):

1. $F_{UN}[N,\rho_N]$ is applicable to any fractional density because the latter can always be considered being asymptotically separated from some other density.
2. $F_{UN}[N,\rho_N]$ can be calculated based on $\rho_N$ only, i.e., without the knowledge of other separated parts. In this sense, a fractional density is an independent entity in the context of DFT.

We also make a few more remarks about $i$-locality. The ANDI condition can always be satisfied in practice without explicit imposition because computers have limited precisions. Therefore, all well-behaved functionals, including the exact and approximated ones, are practically $i$-local. Indeed, LDA (local density approximation) and GGA (generalized gradient approximation) functionals are more local as they are linearly decomposable for two effectively separated densities:

$$
F[N_X+N_Y,\rho^{(X).(Y)}_{N_X.N_Y}]=F[N_X,\rho^{(X)}_{N_X}]+F[N_Y,\rho^{(Y)}_{N_Y}] \tag{7}
$$

We call this type of locality $d$-local ('$d$' stands for density.). A $d$-local functional is also $i$-local. The reverse is not valid in general, e.g., $F_{UN}$ is not $d$-local. Eqs.(3) and (7) also show that an approximate $i$- or $d$-local functional designed for integer charge can be formally extended to the case of fractional charge and maintain their respective locality by simply plugging in a fractional density. On the other hand, a functional that is nonlocal (not $i$-local) must contain nonvanishing energetic interaction between asymptotically separated densities.

## Density functional treatment of the size consistency problem

The $i$-locality concept is also applicable to nuclear external potentials since such a potential can be considered zero numerically for a point placed beyond the AIND from a nucleus, i.e., the *nuclear potential of a molecule may be effectively separable*. Let a spatial domain of a locale with the range defined by the AIND be called *$i$-domain* [Definition D6]. The $i$-domains of a density and its corresponding external potential should coincide for a bound state. Indeed, such an association occurs "naturally" in the constrained search for a lone molecule (the only molecule in the universe) in an asymptotic sense. We observe that the density of such a molecule decays exponentially far away from the molecule, much more rapidly than the external potential from the nuclei[29]. Thus, it is sufficient to constrain the search for its density within a domain limited by the reach of the external potential of the molecule in the two-step Levy constrained search[26]. To formulate it, let $X_L$ symbolize a molecule of $L$ electrons with a set of nuclei $X$. Its electronic energy has the following asymptotic solution:

$$E[L,v_X]=\min_{\rho_L}(F_{UN}[L,\rho_L]+\int\rho_L v_X)=\min_{\rho_L^{(X)}}(F_{UN}[L,\rho_L^{(X)}]+\int\rho_L^{(X)}v_X) \tag{8}$$

where $v_X$ is the external nuclear potential. This energy $E[L,v_X]$ is a functional of $v_X$ for each $L$ and a Legendre transform of $F_{UN}$. The second '$=$' is to say that it is sufficient to search for a density in the $i$-domain of $X$ since the value of $v_X$ is considered zero outside of it. In other words, the integration of $\rho_L^{(X)}v_X$ outside the $i$-domain of $X$ is assumed to be effectively zero because $\rho_L^{(X)}$ is. This relation between the density and external potential is called *$i$-locality association* [Definition D7] henceforth. The minimizer of the search for $E[L,v_X]$ is noted as $\rho_L^X$, the density corresponding to the source of $v_X$. It is customarily called $v$-representable density as it is derived from an external potential. Obviously, the $i$-domain concept is extendable to a subset of nuclei of a molecule when other nuclei are positioned outside of the $i$-domain of the said subset. Note that the *$i$-locality association* is only valid for bound states.

The concept of $i$-locality association can be used to narrow the range of the index $i$ in Eq.(5) in conjunction with a known assumption about the energy of a molecule. The assumption states that the energy of an atom or molecule is strictly convex with respect to the integer number of electrons [3, 30]. We call it *molecular energy convexity condition* [Condition C2] henceforth. Here, we further assume that this condition holds for the energy of a molecule computed in its own $i$-domain and $F_{UN}[L,\rho_L]$ is then convex with respect to $L$ for electrons (Theorem 4.1 in ref[27]). The choice of the index $i$ in Eq.(5) becomes just $\lfloor N\rfloor$ and $\lceil N\rceil$, and $\delta_i$ is fixed by $N$ consequently:

$$F_{UN}[N,\rho_N]=\min_{(1-\delta)\rho_{\lceil N\rceil}+\delta\rho_{\lfloor N\rfloor}=\rho_N}((1-\delta)F_{UN}[\lceil N\rceil,\rho_{\lceil N\rceil}]+\delta F_{UN}[\lfloor N\rfloor,\rho_{\lfloor N\rfloor}]),\quad \delta\equiv\lceil N\rceil-N \tag{9}$$

This form of $F_{UN}[N,\rho_N]$ is the same as derived from the statistical mechanical GCE [3, 16, 31] and the molecular aggregation approaches[17]. Those previous derivations were initiated from particular setups of molecular systems. The derivation here is based on the asymptotic separability of electron density. It admits only those fractional densities that are part of a lone system of integer charge and, thus, is inherently part of $F_{UN}$ for integer numbers of electrons (Lieb's $F_{DM}$ in ref.[27]). In contrast, the GCE formalism[3, 16, 28, 31] extends the definition of $F_{UN}$ beyond Lieb's $F_{DM}$ by admitting mixtures of lone systems with different integer numbers of electrons.

We note that Eq.(8) is consistent with the implementation of DFT on computers in the sense that all computers have limited precisions. However, it contradicts the Hohenberg-Kohn(HK) first theorem [1] in an absolute sense. The theorem establishes a one-to-one correspondence between the density of a nondegenerate molecule and its external potential in an infinite space. As an example of the contradiction, the density of a lone H atom does not change according to Eq.(8) when a positive point charge smaller than 1 is placed outside of the $i$-domain of the H atom. On the other hand, the density must change according to the HK theorem, however small the change might be. This example shows that the result of Eq.(8) can be arbitrarily close to the exact one but is not faithful to the HK theorem nonetheless. Indeed, the theorem requires the density to be above zero anywhere in space[28] and thus infinite precision in implementation, which is nearly impossible in practice. A much worse situation is when the point charge is larger than 1. In this case, the HK theorem would correctly predict the transfer of the electron to the point charge, whereas Eq.(8) would omit this transfer. We show that Eq.(8) and its derivative Eq.(9) for fractional charge can be used to realize the HK theorem for any finite accuracy with a modification of the constraint search in conjunction with the concept of $i$-locality association.

Let $(X..Y)_L$ symbolize a molecule of $L$ electrons with two sets of asymptotically separated nuclei $X$ and $Y$, i.e., $X$ and $Y$ are in separate $i$-domains. Let $E_L^{X..Y}$ be the energy of $(X..Y)_L$ computed with $F_{UN}$. Because the density of this molecule is distributed around $X$ and $Y$ due to the $i$-locality association, it is sufficient to search over trial densities in those two separated $i$-domains. The $i$-local $F_{UN}$ can be applied in each locale, but the outer loop of the Levy's two-step search needs to be divided into two sub-steps, with the outer sub-step being the search of the number of electrons in each $i$-domain, which can be noninteger, in order to cover all possible densities ( $E_{nuc}$ below indicates nuclear repulsion energy.):

$$
\begin{aligned}
& E_L^{X..Y} = \min_{\rho_L}(F_{UN}[L,\rho_L] + \int \rho_L(v_X + v_Y)) + E_{nuc}^{X..Y} \\
& \cong \min_N \min_{\rho_{N..(L-N)}^{(X)..(Y)}} (F_{UN}[L,\rho_{N..(L-N)}^{(X)..(Y)}] + \int (\rho_N^{(X)} + \rho_{L-N}^{(Y)})(v_X + v_Y)) + E_{nuc}^{X..Y} \\
& \cong \min_N \min_{\rho_{N..(L-N)}^{(X)..(Y)}} (F_{UN}[L,\rho_{N..(L-N)}^{(X)..(Y)}] + \int \rho_N^{(X)} v_X + \int \rho_{L-N}^{(Y)} v_Y) + E_{nuc}^X + E_{nuc}^Y \\
& = \min_N (E_N^X + E_{L-N}^Y) = E_{N^*}^X + E_{L-N^*}^Y
\end{aligned}
\tag{10}
$$

with

$$
E_N^X \equiv E[N, v_X] + E_{nuc}^X, \quad E[N, v_X] \equiv \min_{\rho_N^{(X)}}(F_{UN}[N,\rho_N^{(X)}] + \int \rho_N^{(X)} v_X) \tag{11}
$$

$N$ is real and $N^*$ marks the minimizer of the outer loop. The second line in Eq.(10) is due to the density separation according to the $i$-domains existing in $X..Y$ and the $i$-locality association, the third line is due to the use of ANDI condition, and the fourth line due to the $i$-locality of $F_{UN}$ (Eq.(3)) and the $i$-locality association (Eq.(8)). $L$ and $N$ are within the ranges such that the ground states of all three molecules ( $X$, $Y$, and $X..Y$ ) stay bound. The above equations are obviously extendable to a system of an integer number of electrons containing many $i$-domains.

The functional $E[N, v_X]$ is a Legendre transform of $F_{UN}$ for a given $N$. The minimizer of the transform, a $v$-representable density, is denoted as $\rho_N^X$. $E_N^X$ and $\rho_N^X$ depend on the set of nuclei $X$ only.

It is logical within DFT to define the combination of $X$ and $N$ as an independently calculatable entity with $E_N^X$ and $\rho_N^X$ as its energy and density. We call it *fractional molecule* [Definition D8] with the symbol $X_N$, extended from $X_L$. A fractional molecule and its density are effectively separated from other molecules in the universe by definition, i.e., its energy and density can be computed independently of other fractional molecules in the universe. Therefore, the integrations contained in all the equations for a fractional molecule can be computed over the whole space. The $i$-locality association simply means that any interaction sourced outside of the $i$-domain is ignored. In other words, no additional boundary conditions are introduced. The electrons are asymptotically bound inside the $i$-domain of a given set of nuclei because of the inherent satisfaction of the $i$-locality association of the original DFT solution as shown in Eq.(8) and its use in Eq.(11). Thus, there is no need to limit the range of a density explicitly, and the concrete value of AIND is inconsequential for the discussions in this paper. Fractional molecules include molecules of integer charge as special cases.

The extended molecule concept allows the number of electrons of a molecule to change continuously. Physically, the extension fits well with the basic physical assumption in molecular research that atoms and molecules are localizable observables. The electronic energy of a fractional molecule, a Legendre transform of the universal functional, represents a local measurement of the energy of a global system shown as a local partial tracing (Eqs.(1), (2), (9) and (11)). We will use 'molecule' and 'fractional molecule' interchangeably henceforth and use the latter to emphasize the number of electrons being continuous.

The energy and density of a fractional molecule have the following piecewise linearity relations according to Eq.(9):

$$\rho_N^X = (1-\delta)\rho_{\lceil N \rceil}^X + \delta\rho_{\lfloor N \rfloor}^X, \quad \delta \equiv \lceil N \rceil - N \tag{12}$$

and

$$E[N, v_X] = (1-\delta)E[\lceil N \rceil, v_X] + \delta E[\lfloor N \rfloor, v_X], \quad E_N^X = E[N, v_X] + E_{nuc}^X. \tag{13}$$

Obviously,

$$F_{UN}[N, \rho_N^X] = E[N, v_X] - \int \rho_N^X v_X \tag{14}$$

The density and energy of a fractional molecule have the same form as derived from the thermodynamic GCE approach, and the energy is piecewise linear with respect to its number of electrons between integers with $-I(X_{\lceil N \rceil})$ ($I$ stands for ionization potential) as the slope. Consequently, $F_{UN}[N, \rho_N^X]$ is piecewise linear because all the terms on the RHS of Eq.(14) are. The piecewise linearity of $E_N^X$ also means that $N^*$ of Eq.(10) is either an integer or a fractional number between two consecutive integers if the molecular energy convexity condition C2 is extended to each of two asymptotically separated molecules. With this condition, $N^*$ satisfies the following physical condition ($A$ stands for electron affinity):

$$I(X_{\lceil N^* \rceil}) \geq A(Y_{\lfloor L-N^* \rfloor}), \quad A(X_{\lfloor N^* \rfloor}) \leq I(Y_{\lceil L-N^* \rceil}) \tag{15}$$

When $N^*$ is found to be integer only, none of the equalities holds, which is a common occurrence. Otherwise, $N^*$ is found to be between two consecutive integers, and one or both equalities hold with $(X..Y)_L$ being degenerate between $\lfloor N^* \rfloor$ and $\lceil N^* \rceil$. The uniqueness of the range of $N^*$ is ensured by the molecular energy convexity (C2). Assuming that arbitrary $I$ and $A$ values may occur, $N^*$ being a noninteger is always a possibility.

The asymptotically accurate property of piecewise linearity is an extremely stringent requirement for an $i$-local functional[4]. One may wonder if it is possible to design an approximate universal functional $\tilde{F}_{UN}[N,\rho_N^X]$ that coincides with the exact $F_{UN}[N,\rho_N^X]$ when $N$ is an integer only and is not piecewise linear otherwise. The answer is No if $\tilde{F}_{UN}[N,\rho_N^X]$ is $i$-local and continuous with respect to $N$ because we show below that such a functional would intersect the piecewise-linear $F_{UN}[N,\rho_N^X]$ at every possible rational $N$. In other words, the piecewise linearity is a necessary property for an approximate $i$-local functional to be broadly accurate.

Let $G[N,\rho_N^X] \equiv \tilde{F}[N,\rho_N^X] - F[N,\rho_N^X]$ with $G[L,\rho_L^X]=0$. Let $P,Q$ be positive integers and $P<Q$. Suppose we have a system made of $Q$ number of fractional molecules of the same set of nuclei $X$ and $QL+P$ total number of electrons. Then, $\tilde{F}_{UN}[QL+P,\rho_{QL+P}^{X..X..}]$ yields the same value as $F_{UN}[QL+P,\rho_{QL+P}^{X..X..}]$. On the other hand, this system is degenerate when the charge of a molecule differs from the charge of another by no more than one. One of the solutions for Eq.(10) is for the system to have $Q$ identical fractional molecules with $N = L + \dfrac{P}{Q}$ number of electrons for each. We may calculate the value of the universal functional of the system as either $Q(F_{UN}[N,\rho_N^X]+G[N,\rho_N^X])$ or $QF_{UN}[N,\rho_N^X]$. Therefore, $G[N,\rho_N^X]=0$. The possible arbitrary closeness between two points of zero value means that $G[N,\rho_N^X]$ is zero everywhere since it is continuous with respect to $N$.

The final result of Eq.(10) yields the so-called molecular size consistency principle [32] when $N^*$ is an integer. The physically motivated principle has long been used to test approximate methods' correctness. It states that the energy of a system made of two asymptotically separated molecules $X_{L_X}..Y_{L_Y}$ can be calculated in an additive manner, i.e., $E_{L_X..L_Y}^{X..Y} = E_{L_X}^X + E_{L_Y}^Y$. At first glance, the principle seems to suggest that the energy of a molecule as a functional of external potential is $i$-local. In reality, it is not because the electronic charges of the two molecular fragments are decided globally with Eq.(10), i.e., $N^*$ is obtained globally in $X_{N^*}..X_{L-N^*}$ even for integer $N^*$. Furthermore, it is done in a nonlinear manner. This shows the nonlocality character of molecular wavefunctions, with which the number of electrons of a local molecule cannot be decided locally. Obviously, the molecular size consistency principle has been applied routinely and correctly without using Eq.(10). The reason is that the charges are 'manually' optimized in equivalence to the equation. The 'manual' optimization can be achieved by using the ionization potential and electron affinity data or by estimating their differences reasonably, following Eq.(15). The inclusion of the size consistency principle in Eq.(10) adds to the physical validity of fractional molecules. Just like molecular fragments of integer charge being treated as individual molecules in the size consistency principle, the fragments of fractional charges are treated as individual fractional molecules in Eq.(10).

The molecular size consistency principle was known to be a nontrivial matter to approximate wavefunction methods[32]. The derivation of Eq.(10) depends on the $i$-locality of the universal functional only and is applicable to any approximated $i$-local functional, demonstrating the power of DFT. In practice, approximated functionals are often size inconsistent because they are of spin densities, i.e., in a form different from $F_{UN}[N,\rho_N]$. Such a functional may change its value with respect to the variation of spin configuration for the same density as characterized by the so-called fractional spin errors[5]. For cases where size consistency is maintained, approximated functionals can still have large errors for the value of the minimizer of Eq.(10) because they typically assume an integer number of electrons around each electron (See discussions below).

Eq.(10) has a simple physical meaning: Two molecules will exchange electrons to lower the total energy, and the minimum energy point determines the electronic charge for each molecule. The equation is obviously extendable to a system composed of many asymptotically separated fractional molecules. Also, the LHS of the equation can be a fractional molecule due to the $i$-locality of $F_{UN}$ for fractional density (Eq.(6)), i.e., the following generally holds:

$$E_N^{X..Y} = \min_M (E_M^X + E_{N-M}^Y) \ , \tag{16}$$

where $M$ is real. All the discussions above are applicable to a system of many fractional molecules.

The fundamental goal of molecular quantum mechanics is to solve for molecular energy. The present solution for $E_L^{X..Y}$ differs from the exact one in two ways: (1) The ANDI condition C1 is an approximation. For instance, it ignores the possible break of the degeneracy of a system[33] with the presence of another system beyond the AIND. (2) An exact electron density is above zero almost everywhere in the universe due to the unique continuation property of an electronic wavefunction[28], i.e., not exactly separable, and the wavefunction of $X..Y$ cannot be factorized exactly into a left-right one as a consequence. The approximate factorization means that the exact $F_{UN}$ can be approached arbitrarily close by increasing the AIND but may not necessarily be reached, nonetheless. Fundamentally, the application of the ANDI condition means that the solution space is effectively limited to an $i$-domain. $F_{UN}$ of Eq.(9) is a local partial trace, as shown by Eq.(1). The Legendre transform of this functional of a local density yields the desired electronic energy at the same locale. On the other hand, the use of this energy in Eq.(10) misses the infinitesimally small interaction between two asymptotically separated fragments. For instance, the energy difference between the two spin states of an infinitely stretched $H_2$ molecule can never be exactly zero.

Let $\rho \leftrightarrow v$ symbolize the one-to-one mapping between a density and a potential other than a constant in the latter. The HK first theorem reveals a $\rho_L \leftrightarrow v_{ext}$ relation for a nondegenerate molecule of integer charge with $v_{ext}$ being the external nuclear potential. This theorem is also applicable to a fractional molecule. To see this, we make a HK-like *nondegeneracy assumption* on a fractional molecule $X_N$ : Molecules $X_{\lceil N \rceil}$ and $X_{\lfloor N \rfloor}$ be nondegenerate [Condition C3]. Then the relation $\rho_N^X \leftrightarrow v_X$ follows from Eq.(12) due to the $\rho_L^X \leftrightarrow v_{ext}^X$ relation in the $i$-domain ($X$) for integer charges $\lceil N \rceil$ and $\lfloor N \rfloor$ (Eq.(8)). It also means that $\rho_N^X$ is nondegenerate.

The $\rho \leftrightarrow v_{ext}$ relation for a nondegenerate fractional molecule shows the asymptotic nearsightedness of DFT. On the other hand, it means that a practical computation of a molecule implies a theoretical uncertainty because the computation is not sensitive to an arbitrary external potential placed outside of the $i$-domain of the said molecule that may cause a change in the number of electrons of that molecule, i.e., a potential violation of the global HK first theorem as discussed above. Eq.(10) shows how the global HK $\rho \leftrightarrow v_{ext}$ relation can be maintained: For a molecule of integer charge made of two fractional molecules, i.e., $X_N..Y_{L-N}$ with $N$ optimized using Eq.(10), the electron density may change abruptly with an electron being transferred from one molecule to the other when the external potential of the system changes slightly passing through a degeneracy. The localizable $F_{UN}$ treats this change with the number of electrons being an explicit parameter, while the shape of the density at each locale is decided by the external potential at the respective locale only (Eq.(11)). At the degeneracy, $\delta \equiv \lceil N \rceil - N$ varies continuously in $[0,1)$ between two fractional molecules while $E_N^X + E_{L-N}^Y$ remains the same. Otherwise, $N$ is an integer only, and $\rho_N \leftrightarrow v_{ext}$ in each $i$-domain leads to the global HK $\rho_L \leftrightarrow v_{ext}$. Overall, Eq.(10) shows the advantage of DFT in that the global solution to the size consistency problem can be decomposed effectively linearly into local solutions.

We note that fractional molecules are assumed to be nondegenerate in the rest of this paper to facilitate the application of the KS scheme unless noticed explicitly otherwise.

## Kohn-Sham scheme for fractional molecules

At the core of the KS scheme is to solve the noninteracting kinetic energy functional of an interacting $v$-representable density of integer charge. The functional is defined as a search over first-order reduced density matrices (1-RDM, denoted as $\gamma$) constrained by a molecular density $\rho_L^X$:

$$T_{NI}[L, \rho_L^X] \equiv \min_{\gamma_{\rho_L^X}} Tr(-\frac{1}{2}\gamma_{\rho_L^X}\nabla^2) \quad (17)$$

The subscript $NI$ stands for noninteracting. The KS scheme is based on the assumption that an interacting $v$-representable density is simultaneously noninteracting $v$-representable, i.e., $\rho_L \to v_{ext} \Rightarrow \rho_L \to v_{eff}$, with $v_{eff}$ as an effective multiplicative potential of a noninteracting system. The original version of the KS scheme, which is used widely in applications, is more restrictive: a nondegenerate interacting wavefunction $v$-representable density is simultaneously nondegenerate noninteracting wavefunction $v$-representable, i.e. $\rho_L \leftrightarrow v_{ext} \Rightarrow \rho_L \leftrightarrow v_{eff}$. In that case, the solution of Eq.(17) is equivalent to an SD, and the corresponding $T_{NI}$ is often written as $T_S$ with $S$ standing for Slater.

To apply the KS scheme to fractional molecules, we need to extend the definition of $T_{NI}$ to the case of fractional charge and prove that $T_{NI}$ is $i$-local for any noninteracting $v$-representable density:

$$T_{NI}[L, \rho_{N..L-N}^{X..Y}] \cong T_{NI}[N, \rho_N^X] + T_{NI}[L-N, \rho_{L-N}^Y] \quad (18)$$

First, it turns out that Eq.(17) is a valid definition of a functional of fractional density in general because an ensemble density can be represented sufficiently by a 1-RDM [28]. Therefore, we may define the noninteracting kinetic energy functional of a $v$-representable fractional density as:

$$T_{NI}[N, \rho_N^X] \equiv \min_{\gamma_{\rho_N^X}} Tr(-\frac{1}{2}\gamma_{\rho_N^X}\nabla^2) \quad (19)$$

Its application to an electron density requires a KS-like assumption: an interacting ensemble $v$-representable density of fractional charge, which has the $\rho_N \leftrightarrow v_{ext}^X$ relation in an $i$-domain $X$, is simultaneously noninteracting ensemble $v$-representable with a one-to-one mapping, i.e., $\rho_N^X \leftrightarrow v_{ext}^X \Rightarrow \rho_N^X \leftrightarrow v_{eff}^{X_N}$, in the same $i$-domain. The minimizer of $T_{NI}$, denoted as $\gamma_N^X$, is unique within its $i$-domain since $\rho_N^X$ is nondegenerate. It is equivalent to the set of orbitals that are the eigen-solutions for the corresponding $v_{eff}$, with the HOMO fractionally occupied following the Aufbau rule[20]. Those orbitals are customarily called KS orbitals. The singleness of the HOMO is due to the uniqueness of $\gamma_N^X$. $T_{NI}[N,\rho_N]$ becomes $T_S[N,\rho_N]$ when $N$ is an integer. $v_{eff}^{X_N}(\mathbf{r})$ tends to a constant as $\mathbf{r}$ goes out of $i$-domain ($X$).

$X_N..Y_{L-N}$, a lone molecule made of two fractional molecules with $N$ already optimized, can have many densities of varying $\delta(\equiv\lceil N\rceil - N)$ corresponding to the same external potential $v_{ext}^{X..Y}$ when it is degenerate, i.e., $\rho_{N..(L-N)}^{X..Y} \rightarrow v_{ext}^{X..Y}$. We observe that such a many-to-one mapping does not exist for the corresponding noninteracting system because if it did, one noninteracting effective potential $v_{eff}^{X_N..Y_{L-N}}$ of $X_N..Y_{L-N}$ would then be able to generate the interacting densities of both $X_{\lfloor N\rfloor}$ and $X_{\lceil N\rceil}$, which is impossible. Likewise, two different $\delta$ values mean an interacting change of density at each locale and must correspond to different $v_{eff}$'s for the whole system. With the nondegeneracy for a given $\delta$, a KS-like assumption means a $\rho_{N..(L-N)}^{X..Y} \leftrightarrow v_{eff}^{X_N..Y_{L-N}}$ relation. Obviously, this one-to-one correspondence holds for a system made of an arbitrary number of nondegenerate fractional molecules. However, it holds only for an interacting density, as the corresponding effective potential may generate many noninteracting densities, as shown below.

The $\rho_{N..(L-N)}^{X..Y} \leftrightarrow v_{eff}^{X_N..Y_{L-N}}$ relation reveals that $v_{eff}^{X_N..Y_{L-N}}$ in the $i$-domains of $X$ and $Y$, respectively, differs from $v_{eff}^{X_N}$ and $v_{eff}^{Y_{L-N}}$, respectively, by no more than a constant due to the extended KS assumption of $\rho \leftrightarrow v_{eff}$ in each $i$-domain. This division of $v_{eff}^{X_N..Y_{L-N}}$ can be used to prove the $i$-locality of $T_{NI}$ (Eq.(18)) as follows.

A trial KS orbital of $X_N..Y_{L-N}$ in general can be written as the sum of two sides since the density is effectively separated:

$$\varphi_i = c_i^X \varphi_i^{(X)} + c_i^Y \varphi_i^{(Y)} \quad , \tag{20}$$

where $\varphi_i^{(X)}$ is normalized in the $i$-domain $(X)$. The value of $c_i^X$ can be determined *a priori* based on the distribution of the charge. This leads to the split of a trial KS 1-RDM formally as follows, with $f_i$ being the normal spin-summed KS occupancy per the Aufbau rule:

$$\begin{aligned} &\gamma_{\rho_{N..(L-N)}^{X..Y}}(\mathbf{r},\mathbf{r}') = \gamma_{\rho_{N..(L-N)}^{X..Y}}^{(X)}(\mathbf{r}_{(X)},\mathbf{r}'_{(X)}) + \gamma_{\rho_{N..(L-N)}^{X..Y}}^{(Y)}(\mathbf{r}_{(Y)},\mathbf{r}'_{(Y)}) + \sum_{i=1}^{\lceil L/2\rceil} f_i c_i^X c_i^Y \varphi_i^{(X)}(\mathbf{r})\varphi_i^{(Y)}(\mathbf{r}'), \\ &\gamma_{\rho_{N..(L-N)}^{X..Y}}^{(X)}(\mathbf{r}_{(X)},\mathbf{r}'_{(X)}) \equiv \sum_{i=1}^{\lceil L/2\rceil} f_i (c_i^X)^2 \varphi_i^{(X)}(\mathbf{r})\varphi_i^{(X)}(\mathbf{r}') \end{aligned} \tag{21}$$

The first two terms in this equation produce $\rho_N^X$ and $\rho_{L-N}^Y$, respectively. The noninteracting kinetic energy functional of $\rho_{N..(L-N)}^{X..Y}$ can be written as follows because (1) $\left\langle \varphi_i^{(X)} \middle| \nabla^2 \middle| \varphi_i^{(Y)} \right\rangle \cong 0$ ; (2) the density constraint is only applicable to $\gamma^{(X)}_{\rho^{X..Y}_{N..(L-N)}}$ and $\gamma^{(Y)}_{\rho^{X..Y}_{N..(L-N)}}$, all due to the effective separation between $\varphi_i^{(X)}$ and $\varphi_i^{(Y)}$ :

$$T_{NI}[L,\rho^{X..Y}_{N..(L-N)}] \cong \min_{\gamma^{(X)}_{\rho^{X..Y}_{N..(L-N)}}} Tr(-\frac{1}{2}\gamma^{(X)}_{\rho^{X..Y}_{N..(L-N)}}\nabla^2) + \min_{\gamma^{(Y)}_{\rho^{X..Y}_{N..(L-N)}}} Tr(-\frac{1}{2}\gamma^{(Y)}_{\rho^{X..Y}_{N..(L-N)}}\nabla^2) \tag{22}$$

$-\frac{1}{2}Tr(\gamma^{(X)}_{\rho^{X..Y}_{N..(L-N)}}\nabla^2)$ is no less than $T_{NI}[N,\rho_N^X]$ since $\gamma^{(X)}_{\rho^{X..Y}_{N..(L-N)}}$ is also constrained by $\rho_N^X$. It is the same as the latter because $v_{eff}^{X_N..Y_{L-N}}$, which determines $\gamma^{(X)}_{\rho^{X..Y}_{N..(L-N)}}$ in the $i$-domain $(X)$, is the same as $v_{eff}^X$ in $(X)$. Thus, Eq.(18) is proven, and $\gamma^{(X)}_{\rho^{X..Y}_{N..(L-N)}}$ can be written as $\gamma_{\rho_N^X}$.

Obviously, $f_i(c_i^X)^2$ is the KS orbital occupancy in the $i$-domain $(X)$. When $N$ is an integer, $c_i^X$ is either one or zero. When $N$ is not an integer, the $c^X$ of a subHOMO (orbitals below the HOMO(s)) is either one or zero and the HOMO energy of $X_N$ must be the same as that of $Y_{L-N}$ since they are part of a HOMO across the two $i$-domains. The HOMOs of the two fractional molecules form a two-dimensional HOMO space for the whole $X_N..Y_{L-N}$ molecule, and different noninteracting densities may be constructed for a given $v_{eff}^{X_N..Y_{L-N}}$ and $L$, i.e., $\rho_L \to v_{eff}^{X_N..Y_{L-N}}$. One of those densities coincides with the interacting density $\rho_{N..(L-N)}^{X..Y}$ with the values of $f$, $c^X$ and $c^Y$ of the corresponding HOMO(s) being determined by the charge distribution between the two $i$-domains. We note that the solution of Eq.(18) does not determine the sign of the third term of Eq.(21), but the signs are inconsequential because they do not affect the energy. We also note the rank reduction of the KS 1-RDM of $X_N..Y_{L-N}$ in each $i$-domain.

The derivation above obviously supports the extension of Eq.(18) to include more nondegenerate fractional molecules on the RHS. In general, there are as many HOMOs as the number of fractional molecules. All those HOMOs must have the same energy because electrons would be forced to redistribute otherwise. This is an example of applying the KS scheme to a degenerate system if the interacting density of the system and the KS effective potential has a one-to-one correspondence. Also $T_{NI}[N,\rho_N^X]$ is $i$-local for a noninteger $N$ because two fractional molecules can always be considered as part of a system of integer charge, i.e.,

$$T_{NI}[N,\rho^{X..Y}_{M..N-M}] \cong T_{NI}[N,\rho_M^X] + T_{NI}[N-M,\rho_{N-M}^Y] \tag{23}$$

One may wonder if the linear decomposition of $T_{NI}$ in Eq.(18) is applicable to a molecule in an $i$-domain that has two effectively but not asymptotically separated fragments since Eq.(22) still holds, i.e., if $T_{NI}$ is $d$-local. The answer is negative within the KS scheme. The KS $\rho \leftrightarrow v_{eff}$ relation is assumed for an interacting $v$-representable density that is asymptotically separated from other densities as shown above. When the density of a fragment is not asymptotically separated from others, it matches the external potential sourced from all the nuclei within its $i$-domain as part of the total density in that domain. In other words, it does not have the $\rho \leftrightarrow v_{ext}$ relation with those nuclei by itself. In practice, such a decomposition may often be found to be a good approximation when some cancellation effects occur.

The definition of $T_{NI}$ is applicable to wavefunction, ensemble, and grand canonical ensemble $v$-representable densities[28]. We note that it is possible to have an alternative definition of kinetic energy functional as $F_{UN}[N,\rho_N]|_{V_{ee}=0}$, i.e., setting $V_{ee}$ to 0. It corresponds to a GCE of noninteracting Lieb kinetic energy functionals[34]. It is larger than or equal to $T_{NI}$ because an ensemble of 1-RDM remains as a 1-RDM. Physically speaking, a GCE of two KS noninteracting reference systems is not the KS noninteracting solution of the GCE of the corresponding interacting systems because their densities are not derivable from the same KS effective potential, as shown above.

### XC (exchange-correlation) functional for fractional molecules

We may formally define the quantum XC functional of the density of a fractional molecule with the extended KS assumption as:

$$F_{XC}[N,\rho_N^X] \equiv F_{UN}[N,\rho_N^X] - T_{NI}[N,\rho_N^X] - E_J[\rho_N^X] \ , \tag{24}$$

where $E_J$ is the classic Coulomb energy. $F_{XC}$ is $i$-local because all the terms on the RHS of the equation are. It is not $d$-local in general (Eq.(7)).

The key to the success of the KS scheme is the wavefunction representability of the density of integer charge due to the one-to-one mapping between the density and the KS noninteracting effective potential, i.e., $\rho_L \leftrightarrow v_{eff}$ leads to $\rho_L \leftrightarrow \Phi_S(L)$ ($\Phi_S$ stands for a SD.). Besides the computational efficiency of such a representation, the uniqueness of the SD allows the extraction of the major portion of the XC functional by using the noninteracting SD as a trial interacting wavefunction of the 0$^{th}$-order approximation. It yields a unique definition of exchange (the so-called exact exchange) [35], which leads to a unique definition for correlation (C). The separation between X and C permits focused development for each term. Such a uniqueness does not exist for a fractional molecule even if it is considered part of a system of integer charge because the density of a fractional molecule is the result of ensembles of potentially many HOMOs formed from many fractional molecules, as shown above. A further complication arises in the spin space where electrons of either spin may occupy different HOMOs of different fractional molecules, and those HOMOs and their ensembles may result in fractional spin occupancies for each fractional molecule. We show here constructively that a noninteracting ensemble $v$-representable fractional density (i.e., the density of a fractional molecule) is simultaneously noninteracting wavefunction $v$-representable in the sense that it is derivable from the latter. Consequently, it allows a unique definition of an XC functional of fractional density through a constrained search in the spin space.

We first establish a reference system of integer charge for a given $v$-representable fractional density. A fractional molecule can obviously be paired with many fractional molecules, with no change to its $F_{UN}$, to form a degenerate $X..Y$ type of molecule of integer charge. Without loss of generality, we construct a molecule $X_N..A_\delta$ ($\delta \equiv \lceil N \rceil - N$) for a given $X_N$ of noninteger $N$ with the specification that $A_\delta$ has a single nucleus with a nuclear charge satisfying $I(A_1) = I(X_{\lceil N \rceil})$. The special atom is named *auxiliary atom of* $X_N$ [Definition D9] and completely derivable from $X_N$. $F_{UN}[\delta,\rho_\delta^A]$ has a trivial one-electron H-atom-like solution. $X_N..A_\delta$ is called *reference molecule* [Definition D10] and denoted as $\tilde{X}_N$ with $\lceil N \rceil$ electrons. It is of two-fold degeneracy due to $I(A_1)=I(X_{\lceil N\rceil})$ and the nondegeneracy condition of $X_N$ (C3).

All the KS subHOMO orbitals of $\tilde{X}_N$ are on $X_N$. The KS HOMO space of $\tilde{X}_N$ is two-dimensional composed of the HOMOs of $X_N$ and $A_\delta$, respectively. Denoting the HOMO occupancies of $X_N$ and $\tilde{X}_N$ as $n^X$ and $N_h$, respectively, we see that $N_h=\lceil n^X \rceil$. When $N_h=1$, a KS SD for $\tilde{X}_N$ can have one of the two following orbitals as the HOMO:

$$\sqrt{n^X}\varphi_h^X \pm \sqrt{1-n^X}\varphi_h^A \tag{25}$$

where $\varphi_h^X$ and $\varphi_h^A$ are the HOMOs of fractional molecules $X_N$ and $A_\delta$, respectively. The fractional spin occupancies for the HOMO of $X_N$, denoted as $n_\alpha^X$ and $n_\beta^X$, respectively, can be formed by ensembles of the two degenerate KS SDs with the HOMOs of $\tilde{X}_N$ occupied by an $\alpha$ and a $\beta$ electrons, respectively.

When $N_h=2$, the two HOMO electrons have opposite spins because $n^X>1$ and either $n_\sigma^X$ may not exceed one. Those two electrons do not need to occupy the same HOMO, which means that the $\alpha$ and $\beta$ occupancies of the HOMO of $X_N$ may be different from each other. A noninteracting SD for $\tilde{X}_N$ can be formed with the following functions as HOMOs:

$$\sqrt{n_\sigma^X}\varphi_h^X \pm \sqrt{(1-n_\sigma^X)}\varphi_h^A, \quad \sigma=(\alpha,\beta) \tag{26}$$

The fact that the SDs represented by the HOMO functions of Eqs.(25) and (26) can always be constructed validates the relation that *a noninteracting ensemble $v$-representable fractional density is noninteracting wavefunction representable*. Furthermore, the KS orbitals of the fractional molecule can be seen as part of the KS orbitals of the reference molecule, which is to say that the minimizing 1-RDM for the noninteracting kinetic energy of the fractional molecule is derivable from the noninteracting wavefunction of the reference molecule. Thus, it is N-representable, a nice property to have. As an example, a proton with ½ electron has an electron density made from an ensemble between 0 and 1-electron states. It can also be viewed as part of a system made of two asymptotically separated protons with total one electron and its density is derivable from the wavefunction of that system.

The SDs formed above may be used as trial interacting wavefunctions to extract a major portion of the XC energy of $\tilde{X}_N$. When $N_h=1$, the energies of the spin ensembles with the SDs formed from Eq.(25) are the same as the energy of the SD with Eq.(25) being the HOMO with $n_\alpha^X=n^X$. When $N_h=2$, different SDs may be formed with Eq.(26) being the HOMO and the mean-field energies of them differ only in the exchange. In general, the mean-field exchange energy per spin $\sigma$ of a SD wavefunction is an explicit functional of the 1-RDM of that spin: $-\langle \gamma_\sigma^2 \rangle_{ee}$, where $\langle f \rangle_{ee}$ stands for $\frac{1}{2}\int f(\mathbf{r},\mathbf{r}')\left|\mathbf{r}-\mathbf{r}'\right|^{-1} d\mathbf{r}d\mathbf{r}'$. The 1-RDM per spin $\sigma$ for a SD of $\tilde{X}_N$ formed above can be written generally as follows (The choice of '$+/-$' is inconsequential due to the asymptotic separation.):

$$\gamma_{\lceil N\rceil,\sigma}^{\tilde{X}_N}=\gamma_{N,\sigma}^X+\gamma_{\delta,\sigma}^A \pm \sqrt{n_\sigma^X n_\sigma^A}\left(\varphi_h^X(\mathbf{r})\varphi_h^A(\mathbf{r}')+\varphi_h^X(\mathbf{r}')\varphi_h^A(\mathbf{r})\right) \ , \tag{27}$$

where $\gamma^{X}_{N,\sigma}$ is a 1-RDM constructed by filling the KS orbitals of $X_N$ with electrons of spin $\sigma$. The square of $\gamma^{\tilde{X}_N}_{\lceil N\rceil,\sigma}$ is as follows because of $\varphi(\mathbf{r})\varphi^{A}_{h}(\mathbf{r})\cong 0$ for any $\varphi(\mathbf{r})$ on the $X$ side:

$$(\gamma^{\tilde{X}_N}_{\lceil N\rceil,\sigma})^2 = (\gamma^{X}_{N,\sigma})^2 + (\gamma^{A}_{\delta,\sigma})^2 + n^{X}_{\sigma} n^{A}_{\sigma}((\varphi^{X}_{h}(\mathbf{r})\varphi^{A}_{h}(\mathbf{r}'))^2 + (\varphi^{X}_{h}(\mathbf{r}')\varphi^{A}_{h}(\mathbf{r}))^2) \tag{28}$$

The exchange energy due to $\gamma^{\tilde{X}_N}_{\lceil N\rceil,\sigma}$ can be shown to be a sum of $X$ and $A$ sides because of $\left\langle (\varphi^{X}_{h}\varphi^{A}_{h})^2 \right\rangle_{ee} \cong 0$ due to the ANDI condition:

$$-\left\langle (\gamma^{\tilde{X}_N}_{\lceil N\rceil,\sigma})^2 \right\rangle_{ee} \cong -\left\langle (\gamma^{X}_{N,\sigma})^2 \right\rangle_{ee} - \left\langle (\gamma^{A}_{\delta,\sigma})^2 \right\rangle_{ee} \tag{29}$$

This motivates the definition of a nominal spin summed exchange energy for a fractional molecule as follows:

$$E^{X_N}_{EX}(n^{X}_{\alpha}, n^{X}_{\beta}) \equiv -\left\langle (\gamma^{X}_{N,\alpha})^2 \right\rangle_{ee} - \left\langle (\gamma^{X}_{N,\beta})^2 \right\rangle_{ee} \tag{30}$$

It can be sufficiently indexed by the HOMO spin occupancies since the subHOMO orbitals are doubly occupied. The following $i$-locality relation holds between the nominal spin summed exchange energy of the reference molecule and those of its components per Eq.(29):

$$E^{\tilde{X}_N}_{EX}(n^{X}_{\alpha}, n^{X}_{\beta}, n^{A}_{\alpha}, n^{A}_{\beta}) = E^{X_N}_{EX}(n^{X}_{\alpha}, n^{X}_{\beta}) + E^{A_\delta}_{EX}(n^{A}_{\alpha}, n^{A}_{\beta}) \tag{31}$$

We note that the choice of the spin HOMO occupancies is limited due to the SDs formed from Eqs.(25) and (26):

$$\begin{aligned} &N_h = 1\,(n^X < 1): \quad n^{X}_{\alpha} = n^X,\, n^{X}_{\beta} = 0,\, n^{A}_{\alpha} = 1 - n^X,\, n^{A}_{\beta} = 0 \\ &N_h = 2\,(n^X > 1): \quad n^{X}_{\beta} = n^X - n^{X}_{\alpha},\, n^{A}_{\alpha} = 1 - n^{X}_{\alpha},\, n^{A}_{\beta} = 1 - n^X + n^{X}_{\alpha} \end{aligned} \tag{32}$$

We see that $E_{EX}$'s for $X_N$, $A_\delta$ and $\tilde{X}_N$ are functionals of their own respective densities for . $n_X < 1$ .because they are uniquely defined for each $n_X$. We name them exchange-correlation of mean-field(XCMF) as they are due to the mean-field effect of the SD representability, i.e.,

$$F_{XCMF}[N, \rho^{X}_{N}] \equiv E^{X_N}_{EX}(n^X, 0), \quad F_{XCMF}[\lceil N\rceil, \rho^{\tilde{X}_N}_{\lceil N\rceil}] \equiv E^{\tilde{X}_N}_{EX}(n^X, 0, 1-n^X, 0) \tag{33}$$

The following $i$-locality relation holds per Eq.(31):

$$F_{XCMF}[\lceil N\rceil, \rho^{\tilde{X}_N}_{\lceil N\rceil}] = F_{XCMF}[N, \rho^{X}_{N}] + F_{XCMF}[\delta, \rho^{A}_{\delta}] \tag{34}$$

When $n_X > 1$, all three $E_{EX}$'s change with $n^{X}_{\alpha}$. We first define the functional $F_{XCMF}$ for $X_N$ and $\tilde{X}_N$, respectively, by searching among those trial solutions indexed by $n^{X}_{\alpha}$:

$$\begin{aligned} &F_{XCMF}[N, \rho^{X}_{N}] \equiv \min_{n^X/2 \le n^{X}_{\alpha} \le 1} E^{X_N}_{EX}(n^{X}_{\alpha}, n^X - n^{X}_{\alpha}) \\ &F_{XCMF}[\lceil N\rceil, \rho^{\tilde{X}_N}_{\lceil N\rceil}] \equiv \min_{n^X/2 \le n^{X}_{\alpha} \le 1} E^{\tilde{X}_N}_{EX}(n^{X}_{\alpha}, n^X - n^{X}_{\alpha}, 1 - n^{X}_{\alpha}, 1 - n^X + n^{X}_{\alpha}) \end{aligned} \tag{35}$$

It can be viewed as maximizing the exchange effect since the latter is negatively defined. Due to its quadratic nature, $E_{EX}^{X_N}$ is minimized when only one occupancy is noninteger[20], i.e., $n_\alpha^X = 1$ is the minimizer for $F_{XCMF}[N, \rho_N^X]$. Furthermore, $n_\alpha^X = 1$ is the minimizer for $F_{XCMF}[\lceil N \rceil, \rho_{\lceil N \rceil}^{\tilde{X}_N}]$ too because it minimizes both terms contained in $E_{EX}^{\tilde{X}_N}$ (See Eq.(31).), i.e.,

$$
\begin{aligned}
&F_{XCMF}[N, \rho_N^X] = E_{EX}^{X_N}(1, n^X - 1), \\
&F_{XCMF}[\lceil N \rceil, \rho_{\lceil N \rceil}^{\tilde{X}_N}] = E_{EX}^{\tilde{X}_N}(1, n^X - 1, 0, 2 - n^X)
\end{aligned}
\tag{36}
$$

The second line in the equation can be further reduced due to the $i$-locality relation Eq.(31):

$$
F_{XCMF}[\lceil N \rceil, \rho_{\lceil N \rceil}^{\tilde{X}_N}] = F_{XCMF}[N, \rho_N^X] + F_{XCMF}[\delta, \rho_\delta^A] \tag{37}
$$

The solution provided by the above two equations can also be understood as the minimization of self-interaction such that an SD with maximumly separated of $\alpha$ and $\beta$ electrons in the HOMO space is preferred. The search space $n^X/2 \le n_\alpha^X \le 1$ is equivalent to a search over all spin 1-RDM's constrained by their sum to be the spin-summed KS 1-RDM. $F_{XCMF}[\lceil N \rceil, \rho_{\lceil N \rceil}^{\tilde{X}_N}]$ is an upper bound to $F_{XC}[\lceil N \rceil, \rho_{\lceil N \rceil}^{\tilde{X}_N}]$ because the trial solutions are a subset of all possible wavefunctions and their ensembles constrained by the density.

We may now write $F_{XCMF}$ of a fractional molecule for the entire range of HOMO occupancy ($0 < n^X \le 2$) as:

$$
F_{XCMF}[N, \rho_N^X] = E_{EX}^{X_N}(\min(1, n^X), n^X - \min(1, n^X)) \tag{38}
$$

We note that $F_{XCMF}$ is $i$-local but not $d$-local because the ANDI condition is used in Eq.(29) for its derivation.

The remaining energy of $F_{XC}$ is due to the interelectronic interaction of many electrons beyond the mean-field effect of the noninteracting 1-RDM. We name it correlation of multi-electron (CME) origin:

$$
F_{CME} \equiv F_{XC} - F_{XCMF} \tag{39}
$$

This definition is applied to both $X_N$ and $\tilde{X}_N$. The following $i$-locality relation holds since $F_{XC}$ is $i$-local:

$$
F_{CME}[\lceil N \rceil, \rho_{\lceil N \rceil}^{\tilde{X}_N}] = F_{CME}[N, \rho_N^X] + F_{CME}[\delta, \rho_\delta^A] \tag{40}
$$

$F_{CME}$ is nonpositively defined for $\tilde{X}_N$, since $F_{XCMF}$ is an upper bound to $F_{XC}$ for $\tilde{X}_N$. $F_{CME}$ for $A_\delta$ is zero because $F_{XC} = F_{FCMF}$ for $N \le 1$. Thus, the CME of a reference molecule is all on the $X$ side. The nonpositivity of $F_{CME}$ shows that $F_{XCMF}$ is a rigorous upper bound to $F_{XC}$ for a fractional molecule.

An alternative route exists to derive $F_{XCMF}$ for a $v$-representable fractional density $\rho_N^X$ (Eq.(38)) directly without using a reference system. One can start from $\gamma_{N,\sigma}^X$ based on the KS 1-RDM, define the nominal $E_{EX}^{X_N}(n_\alpha^X, n_\beta^X)$ as in Eq.(30), and define $F_{XCMF}[N, \rho_N^X]$ in Eq.(35). The search in Eq.(35) for $n^X \le 1$ reaches $n_\alpha^X = n^X$ as the minimizer as well due to the quadratic nature of the exchange formula, resulting in Eq.(38). We may call the existence of the minimizer constrained by the KS 1-RDM as *Maximum Exchange Principle*, i.e.,

$$F_{XCMF}[N, \rho_N^X] = \min_{\gamma_\alpha + \gamma_\beta = \gamma_N^X} (-\langle(\gamma_\alpha)^2\rangle_{ee} - \langle(\gamma_\beta)^2\rangle_{ee}) \tag{41}$$

The principle can be used to define $F_{XCMF}$ for an ensemble of many nondegenerate fractional molecules with a total integer charge since the $i$-locality decomposition of the nominal exchange (Eq.(29)) obviously holds. Thus, minimizing the total exchange energy becomes minimizing the exchange energy for each fractional molecule. Likewise, it can be used to define $F_{XCMF}$ for an ensemble of many nondegenerate fractional molecules with a total noninteger charge because such an ensemble can always be made as part of a system of integer charge. This shows that $F_{XCMF}$ is $i$-local in general. Still, a reference system of integer charge is needed to prove that $F_{XCMF}$ is an upper bound to $F_{XC}$.

Eq. (38) shows that the derivative of $F_{XCMF}$ with respect to $N$ is discontinuous when the HOMO occupancy passes through one. This reflects the quantum effect of electron pairing of two asymptotically separated electrons in the context of KS noninteracting reference, as the HOMO occupancy of the reference molecule changes from 1 to 2. Adding an electron in the HOMO orbital incurs the Coulomb interaction from the one already there. Indeed, one may find with some algebraic manipulation that the derivative of $F_{XCMF}$ with respect to $n^X$ upshifts by $\langle(\varphi_h^X(\mathbf{r})\varphi_h^X(\mathbf{r}'))^2\rangle_{ee}$, the Coulomb energy between the two electrons, as $n^X$ increases through one by an infinitesimal amount. This observation is consistent with the argument of Perdew et al. based on the GCE treatment that the KS XC potential upshifts as the number of electrons is increased by an infinitesimal amount from an integer[3]. The Coulombic repulsive nature is also consistent with the popular Hubbard model for strongly correlated materials [36].

While $F_{XCMF}[N, \rho_N^X]$ is derivable from an SD wavefunction, it is a local partial trace of the SD with the global Hamiltonian when $N$ is a noninteger and does not possess all the properties of the SD. A well-applied property of an SD is that its exchange hole function at a point in space $\mathbf{r}$ (reference point), defined as $-\rho_\sigma(\mathbf{r})^{-1}\gamma_\sigma^2(\mathbf{r}, \mathbf{r}')$, integrates to -1 at any point in space over $\mathbf{r}'$ (or the size of the exchange hole size is said to be 1). This is known as the exchange sum rule. The correlation hole, on the other hand, integrates to zero. Those sum rules are equivalent to the assumption that there are an integer number of electrons of the same spin around each electron and are known to be inapplicable to a system of noninteger charge [37]. Indeed, one can easily see that the formal exchange hole functions for the computation of $F_{XCMF}[N, \rho_N^X]$ (the RHS of Eq.(38)) integrates to a value between 0 and -1 from the following equality:

$$1 - \rho_{N,\sigma}^X(\mathbf{r})^{-1} \int (\gamma_{N,\sigma}^X(\mathbf{r}, \mathbf{r}'))^2 d\mathbf{r}' = (n_\sigma^X - (n_\sigma^X)^2)(\varphi_h^X(\mathbf{r}))^2 \rho_{N,\sigma}^X(\mathbf{r})^{-1} \tag{42}$$

They integrate to -1 only when $N$ equals 1 or 2. On the other hand, the sum rules are applicable to the reference molecule.

The KS 1-RDM of a fractional molecule does not have a unique spin configuration in general that can be used to define an exact exchange except for the cases of integer spin occupancies where $F_{XCMF}$ becomes the KS exact exchange. However, in situations where such uniqueness can be asserted, $F_{XCMF}$ can be used to define a correlation. One example often discussed in the literature is the dissociation of an $H_2$ molecule, a case of strong correlation. The spin configuration of an H atom in this system can be asserted as $n_\alpha = n_\beta = \frac{1}{2}$, and the correlation then becomes $F_{XCMF}[1, \rho_1^H] - E_{EX}^{H_1}(\frac{1}{2}, \frac{1}{2})$ $= -\frac{1}{2}\left\langle (\varphi_h(\mathbf{r})\varphi_h(\mathbf{r}'))^2 \right\rangle_{ee}$ ($F_{CME}$ is zero). A related finding in the literature was that an accurate functional should remain constant with respect to the variation of fractional spin occupancies in the HOMOs[5]. $F_{XCMF}$ trivially satisfy this condition since it is determined by the total density only. It becomes the KS exact exchange when the HOMO occupancies of both spins are integers.

**Examples of fractional charge**

The challenge of the fractional charge problem is often illustrated with systems of a few electrons. One well-designed example[7] is a cluster of eight geometrically equivalent H nuclei placed very far from each other with the number of electrons ranging from zero to two per H. Full configuration interaction calculations of this system with a minimum basis set showed that the energy surface is flat with respect to the variations of number of electrons and their spins, with a straight ridge along the line of eight electrons (one per H). This flat-plane condition is obvious but not satisfied by commonly used approximate functionals.

We may treat this cluster as eight equivalent effectively independent fractional H atoms with each H modeled with $(n_\alpha, n_\beta)$. H and $H^-$ obviously satisfy the nondegeneracy condition (C3). One can see that $F_{XCMF}[n_\alpha + n_\beta, \rho_{n_\alpha + n_\beta}^H]$ (Eq.(38)) applied to the so-constructed fractional H atom satisfies this flat-plane condition trivially. The energy surface with respect to $(n_\alpha, n_\beta)$ is flat since the orbital does not change with the minimum basis and $F_{XCMF}[n_\alpha + n_\beta, \rho_{n_\alpha + n_\beta}^H]$ is a function of $n_\alpha + n_\beta$. When $n_\alpha + n_\beta \le 1$, $E_J + F_{XCMF} = 0$. When $n_\alpha + n_\beta > 1$, $E_J + F_{XCMF}$ becomes the Coulomb energy between one $\alpha$ electron and $(n-1)$ $\beta$ electron, causing a ridge at $n = 1$. This sudden appearance of the Coulomb interaction is the driving force for the derivative discontinuity at $n = 1$ due to the addition of an electron as discussed above. (Note that $F_{CME}$ (dynamic correlation) is suppressed since the orbital stays the same for one and two electrons with the minimum basis.) Fig. 1 shows the errors with several contemporary functionals along the line of $(\frac{1}{2}n, \frac{1}{2}n)$. The errors for contemporary functionals such as B3LYP[38] and PBE exchange (PBE-X) [39] were well documented in the literature. They are mainly caused by self-interaction and the use of the conventional XC sum-rules. The relatively recent KP16/B13[10] yielded the exact result for $(\frac{1}{2}, \frac{1}{2})$ for a single H, representing an advance in DFT in treating strong correlation. Still, it fails when $n>1$. The reason is that it applies the standard XC sum rules for XC. While it is correct for $n=1$, it is incorrect for $n>1$ as discussed above.

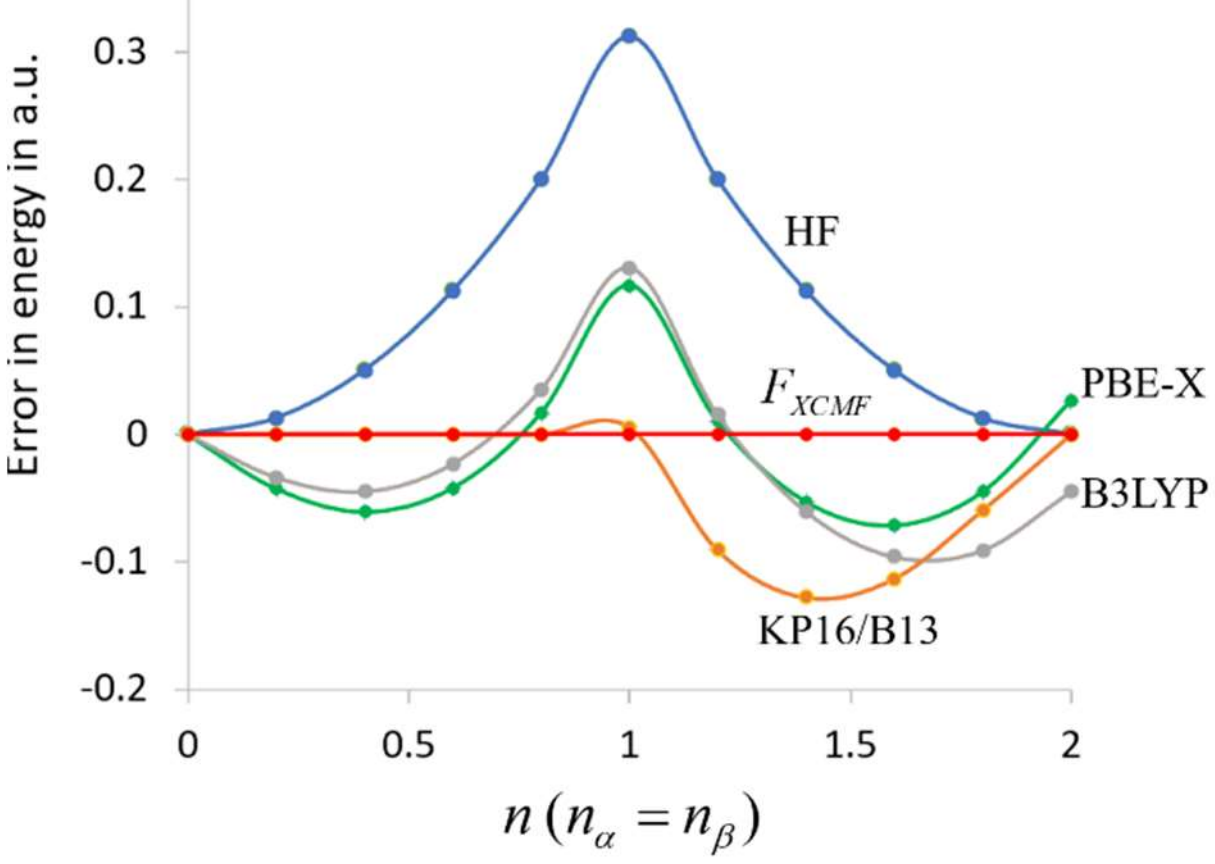

Figure 1. Comparison of various DFT functionals for H atom of fractional occupancies with a minimum basis.

Fig. 2 shows the DFT components of the energy with a better approximated fractional density for $1\leq n\leq 2$. The system is an ensemble of the ground states of $He^+$ and He, with the densities of the two states calculated with Hartree-Fock (HF) plus the Becke-Roussel'94 (BR94) correlation functional[40] with a large basis set. The data shown are relative to the straight line between the two end points ( $n$ at 1 and 2) for each component. The convexity of the Hartree and kinetic energies and the concavity of the XC energy for fractional charge are expected. These two opposite effects largely cancel each other, and their total sum (red line) is somewhat concave, showing that the dynamic correlation $F_{CME}$ needs to be convex in order to achieve linearity. Just as a reference point, the green line shows the correlation energy calculated with BR94. It has the desired convexity although it does not fully compensate the red line. Note that the calculations for Figs 1 and 2 were done using xTron, an in-house DFT program[41].

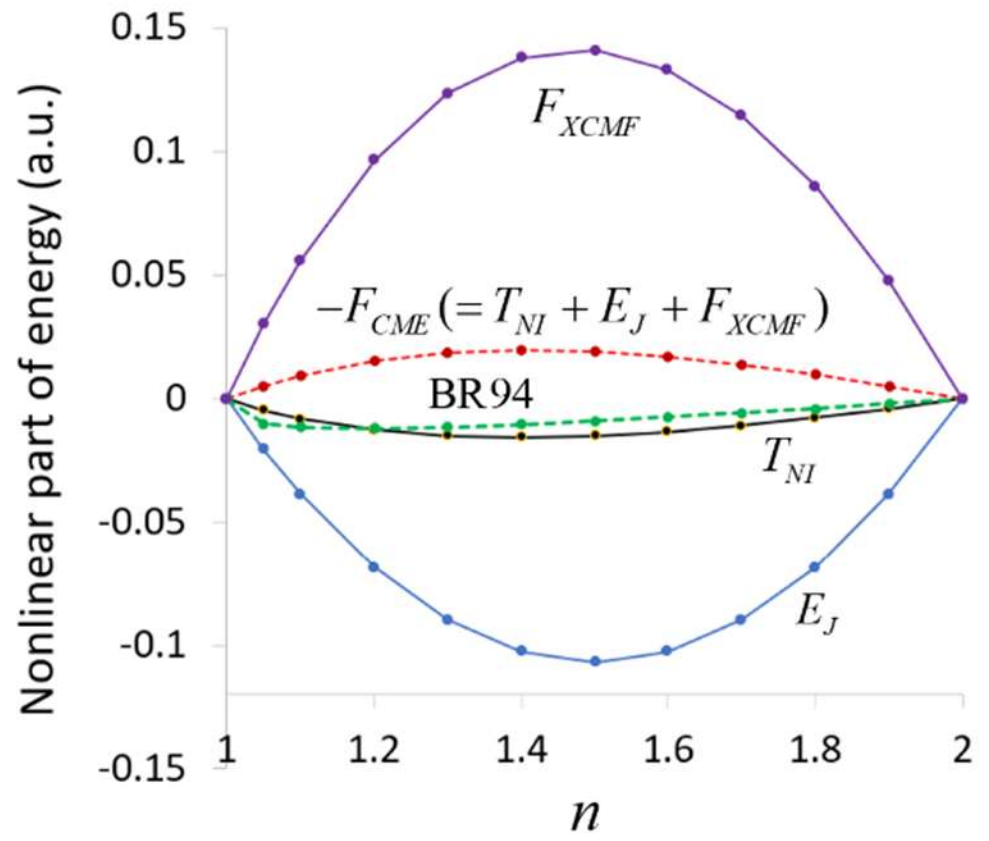

Figure 2. The components of the DFT energy for He atom with fractional occupancy($n$) from 1 to 2.

**Conclusions**

We show that the exact universal density functional of integer electronic charge leads to an extension to fractional charge in an asymptotic sense when it is applied to a system made of asymptotically separated densities. The extended functional is asymptotically local and is said to be $i$-local. The concept of $i$-locality is also applicable to nuclear external potentials, and a natural association exists between the localities of a density and a set of nuclei. Applying the functional to a system with nuclei distributed in two asymptotically separated locales requires an explicit search of the electronic charge at each locale with the constraint of the global charge. The determined number of electrons at each locale can be fractional. The molecular size consistency principle is realized as the result of the search. It is physically sensible to extend the molecule concept to include a fractional number of electrons as a localized observable. The physical validity of fractional molecules is equivalent to the asymptotic separability of molecules, a basic assumption in molecular research. A one-to-one mapping between the density and the external potential of a fractional molecule is shown to exist with a nondegenerate condition. The global one-to-one mapping required by the HK first theorem is realized through the aforementioned global search for molecular charges. Furthermore, the well-known piecewise linearity of the functional with respect to the number of electrons is necessary for an approximate $i$ -local universal functional to be broadly accurate for integer numbers of electrons. The KS noninteracting kinetic energy functional for a fractional molecule is well defined and has the same form as that for a system of an integer number of electrons. It is shown to be $i$ -local. A nondegenerate noninteracting ensemble $v$ -representable fractional density is simultaneously noninteracting wavefunction representable. A constrained search over those representing wavefunctions yields the definition of an exchange-correlation functional pertaining to fractional occupancies. The functional is shown to be an upper bound to the formal Kohn-Sham exchange-correlation energy of a fractional molecule and includes strong correlation. It yields the correct result for a well-designed example of effective fractional occupancies in the literature.

Acknowledgement:

The author expresses deep gratitude to his long-time collaborator, Dr. Emil Proynov, for extensive comments of this manuscript. The paper has gone through major revisions in arXiv Versions v3 and v4 after discussions with Dr. E. J. Baerends. Discussions with Drs. B. G. Janesko, D. Ye, H. Hu and Mr. C. Meherg are also appreciated for arXiv Versions v6 and v7. The author appreciates a reviewer's suggestion of changing the title of the paper. This work received support from the National Science Foundation (Grant No. 1665344).

## Appendix

We follow the bipartite partition of a quantum state that is composed of two wavefunctions that belong to two spatially separated Hamiltonians, a widely accepted practice for the study of quantum entanglement. This type of wavefunction is also often used in quantum chemistry literature for describing the so-called left-right correlation, a strong correlation between two separated electrons. Here, we call this type of wavefunction left-right (L-R) wavefunction. Thus, a wavefunction producing the density $\rho^{(X)..(Y)}_{L_X..L_Y}$, noted as $\Psi^{\cdot\cdot}$, can be expressed as follows:

$$\Psi^{\cdot\cdot}_{\rho^{(X)..(Y)}_{L_X..L_Y}}(L_X+L_Y) \equiv \mathscr{A}[\Psi_{\rho^{(X)}_{L_X}}(L_X)\Psi_{\rho^{(Y)}_{L_Y}}(L_Y)] \tag{43}$$

The contraction of $\Psi^{\cdot\cdot}_{\rho^{(X)..(Y)}_{L_X..L_Y}}$ with the universal part of a Hamiltonian has the following expression:

$$U[\Psi^{\cdot\cdot}_{\rho^{(X)..(Y)}_{L_X..L_Y}}] \equiv \left\langle \Psi^{\cdot\cdot}_{\rho^{(X)..(Y)}_{L_X..L_Y}} \middle| T+V_{ee} \middle| \Psi^{\cdot\cdot}_{\rho^{(X)..(Y)}_{L_X..L_Y}} \right\rangle \overset{ANDI}{\cong} U[\Psi_{\rho^{(X)}_{L_X}}(L_X)] + U[\Psi_{\rho^{(Y)}_{L_Y}}(L_Y)] \tag{44}$$

Electrons can also be in a mixed state, described by a density matrix:

$$D(L) = \sum_i d_i^D \left|\Psi_i^D(L)\right\rangle\left\langle\Psi_i^D(L)\right|, \quad d_i^D \geq 0, \quad \sum_i d_i^D = 1 \tag{45}$$

The universal part of the energy for $D(L)$ is:

$$U[D(L)] = \sum_i d_i^D U[\Psi_i^D(L)] \tag{46}$$

We denote a density matrix producing the density $\rho_L$ as $D_{\rho_L}(L)$. A density matrix producing the density $\rho^{(X)..(Y)}_{L_X..L_Y}$ can be expressed as follows:

$$D^{\cdot\cdot}_{\rho^{(X)..(Y)}_{L_X..L_Y}}(L_X+L_Y) = \sum_{i,j} d_i^{D_{\rho^{(X)}_{L_X}}} d_j^{D_{\rho^{(Y)}_{L_Y}}} \left| \mathscr{A}[\Psi_i^{D_{\rho^{(X)}_{L_X}}}(L_X)\Psi_j^{D_{\rho^{(Y)}_{L_Y}}}\Psi(L_Y)] \right\rangle \left\langle \mathscr{A}[\Psi_i^{D_{\rho^{(X)}_{L_X}}}(L_X)\Psi_j^{D_{\rho^{(Y)}_{L_Y}}}\Psi(L_Y)] \right| \tag{47}$$

The density matrix is named as the L-R type in consistency with the L-R wavefunctions. The universal part of the energy for $D^{\cdot\cdot}_{\rho^{(X)..(Y)}_{L_X..L_Y}}(L_X+L_Y)$ is the following based on Eqs.(44) to (47):

$$U[D^{\cdot\cdot}_{\rho^{(X)..(Y)}_{L_X..L_Y}}(L_X+L_Y)] \overset{ANDI}{\cong} \sum_{i,j} d_i^{D_{\rho^{(X)}_{L_X}}} d_j^{D_{\rho^{(Y)}_{L_Y}}} U[\Psi_i^{D_{\rho^{(X)}_{L_X}}}(L_X)] + \sum_{i,j} d_i^{D_{\rho^{(X)}_{L_X}}} d_j^{D_{\rho^{(Y)}_{L_Y}}} U[\Psi_j^{D_{\rho^{(Y)}_{L_Y}}}(L_Y)] \tag{48}$$

which leads to Eq.(1). Eq.(48) is more general than Eq.(44) since a wavefunction can form a density matrix.

We note that the above derivation can be trivially extended to a system composed of more than two asymptotically separated densities such that the linear decomposition of Eq. (1) remains. As an example, Eq.(50) below for three asymptotically separated densities of integer charge can be derived from Eq.(49) following the same process as from Eq.(43) to Eq.(48) above:

$$\Psi^{\cdot\cdot}_{\rho^{(X1)..(X2)..(X3)}_{L_{X1}..L_{X2}..L_{X3}}}(L_{X1}+L_{X2}+L_{X3}) \equiv \mathscr{A}[\Psi_{\rho^{(X1)}_{L_{X1}}}(L_{X1})\Psi_{\rho^{(X2)}_{L2}}(L_{X2})\Psi_{\rho^{(X3)}_{L_{X3}}}(L_{X3})] \tag{49}$$

$$U[D_{\rho_{L_{X1}..L_{X2}..L_{X3}}^{(X1)..(X2)..(X3)}}(L_{X1}+L_{X2}+L_{X3})] \overset{ANDI}{\cong} U[D_{\rho_{L_{X1}}^{(X1)}}(L_{X1})]+U[D_{\rho_{L_{X2}}^{(X2)}}(L_{X2})]+U[D_{\rho_{L_{X3}}^{(X3)}}(L_{X3})] \qquad (50)$$